\documentclass[12pt, preprint]{aastex}
\begin{document}
\title{What Magnetar Seismology can Teach us about the Magnetic
Fields}\author{Rashid Shaisultanov\altaffilmark{2}, David Eichler \altaffilmark{1}} \altaffiltext{1}{Physics
Department, Ben-Gurion University, Be'er-Sheva 84105, Israel;
eichler@bgu.ac.il} \altaffiltext{2}{Physics Department, Ben-Gurion
University, Be'er-Sheva 84105, Israel; rashids@bgu.ac.il}
\begin{abstract}

   The effect of magnetic fields  on the  frequencies of toroidal oscillations of neutron
stars is derived to lowest order. Interpreting the fine structure in
the QPO power spectrum of magnetars following giant flares reported
by Strohmayer and Watts (2006) to be "Zeeman splitting" of
degenerate toroidal modes, we estimate a crustal magnetic field  of
order $10^{15}$ Gauss or more. We suggest that residual m, -m
symmetry following such splitting might allow  beating of individual
frequency components that is slow enough to be observed.

\end{abstract}

The discovery of quasi-periodic oscillations (QPO) in the
hyperflares of soft gamma-ray repeaters \citep{isr, str05, wat06,
str06} has attracted much attention to the study of nonradial
oscillations of neutron stars with solid crust. These oscillations
were studied extensively in \citep{hans, mcd} for non-magnetic case.
\citet{unnob} discuss in their book many aspects of the theory of
nonradial oscillations of stars. The properties of nonradial modes
of strongly magnetized neutron stars have been investigated by
several authors \citep{dun98, pir05, lee1, lee2, sot1, sot2}. Their
main focus was on the study of axisymmetric modes. Duncan
anticipated toroidal oscillations as a result of giant flares and
noted that the magnetic field could affect their frequencies. The
effects of a strong vertical magnetic field on the oscillation
spectrum of a cylindrical slab model were studied in \citep{car86}.
The nonradial modes are generally divided into two main classes: the
spheroidal and toroidal modes. We will concentrate on the study of
toroidal modes because of their possible connection with QPOs. They
are defined by conditions

\begin{equation}
\vec{\protect\nabla}\cdot \vec{u} =0\ ,u_{r}=0
\end{equation}
where $\vec{u}$ is a displacement vector.

Without magnetic field and rotation, toroidal modes ( denoted $
_{l}t_{n} $, where the index n is the number of radial nodes in the
eigenfunction ) have frequencies that do not depend on m. This
degeneracy is lifted by magnetic field since it breaks the spherical
symmetry of the problem. In this paper we will study how this
happens considering the magnetic field $\vec{B}$ as a perturbation.
The influence of a
 magnetic perturbation on
spheroidal modes was considered earlier [see e.g. \citep{unnob}]. We
also refer to \citet{dah} for a detailed discussion of perturbation
theory with applications in seismology. Assuming an oscillatory time
dependence $\vec{u}\propto e^{-i \omega t}$, where $\omega$ is the
mode frequency, the equation for eigenfunctions without a magnetic
field is
\begin{equation}
-\omega ^{2}\vec{u}=A(\vec{u})
\end{equation}
where the linear operator A describes the dynamics given specific
parameters for the neutron star. We do not need its
exact form here. Interested readers may consult e.g. \citep{hans, mcd}.
Let us denote toroidal eigenfunctions and frequencies obtained from
this equation as $\vec{u}_{lm}^{(0)}$ and $\omega_{l}^{(0)}$. In
spherical coordinates components of $\vec{u}_{lm}^{(0)}$ are
\begin{eqnarray}
u_{\protect\theta }^{(0)} =\frac{w_{l}^{(0)}\left( r\right) }
{\sin \protect\theta }\frac{%
\partial Y_{lm}}{\partial \protect\varphi }\ ,u_{\protect\varphi }^{(0)}=-w_{l}^{(0)}\left(
r\right) \frac{\partial Y_{lm}}{\partial \protect\theta }.
\protect\end{eqnarray}
Here $w_{l}^{(0)}\left( r\right)$ is a radial eigenfunction.

 With a magnetic field we have

\begin{equation}\label{a4}
-\omega ^{2}\vec{u}=A(\vec{u})+\frac{1}{4\pi \rho }[(\vec{\nabla}\times \vec{%
b})\times \vec{B}]
\end{equation}
where
\begin{equation}\label{a5}
\vec{b}=\vec{\nabla}\times (\vec{u}\times \overrightarrow{B})
\end{equation}
We now take $\vec{B}$ to be the uniform field
\begin{equation}
\vec{B}=B_{0}\vec{e}_{z}.
\end{equation}
It is helpful to use vector spherical harmonics [see e.g.
\citep{var}] to express the vector operators in equations (\ref{a4})
and (\ref{a5}) in $Y_{lm}$ representation. Definitions of vector
spherical harmonics and some useful formulae are given in the
appendix. Then, using the perturbation approach described in
\citet{unnob}, we obtain
\begin{equation}
\omega _{lm}  = \omega _{l}^{(0)}+\omega _{lm}^{(1)}
\end{equation}
where
\begin{eqnarray}
\frac{\omega _{lm}^{(1)}}{\omega _{l}^{(0)}} &=&-\frac{B_{0}^{2}}{8\pi   \label{a1}
\omega _{l}^{(0)2}}\frac{1}{\int \rho w_{l}^{(0)2}r^{2}dr}\int r^{2}drw_{l}^{(0)}\left[ \frac{%
1}{r^{2}}\frac{d}{dr}\left( r^{2}\frac{dw_{l}^{(0)}}{dr}\right) -\frac{l\left(
l+1\right) }{r^{2}}w_{l}^{(0)}\right] F\left( l,m\right),\label{a}
\end{eqnarray}
where
\begin{equation}\label{b1}
F\left( l,m\right) =\frac{l\left( l+2\right) \left( \left(
l+1\right) ^{2}-m^{2}\right) }{\left( l+1\right) ^{2}\left(
2l+1\right) \left( 2l+3\right) }+\frac{m^{2}}{l^{2}\left( l+1\right)
^{2}}+\frac{\left( l^{2}-1\right) \left( l^{2}-m^{2}\right)
}{l^{2}\left( 2l+1\right) \left( 2l-1\right) }\label{b}
\end{equation}
Table 1 lists values of $F\left( l,m\right)$ for specific modes.

This is the main result of our paper within the framework of a
perturbative calculation. Perturbation theory is applicable when
\begin{equation}\label{c9}
\frac{\omega _{lm}^{(1)}}{\omega _{l}^{(0)}}\ll 1
\end{equation}
For the case of a spherical star with uniform shear modulus $\mu$,
one can
calculate the integral in (\ref{a1}) exactly and obtain
\begin{equation}
\frac{\omega _{lm}^{(1)}}{\omega _{l}^{(0)}}=\frac{B_{0}^{2}}{8\pi \mu }%
F\left( l,m\right)
\end{equation}
 By (\ref{c9}) perturbation theory is thus applicable when
\begin{equation}
\frac{B_{0}^{2}}{8\pi \mu }\ll 1
\end{equation}
Because of the axisymmetry of the magnetic field the (2l+1)
degeneracy in m is only partially lifted : it depends only on $|m|$.
The mode splits into (l+1) modes.

The "92 hertz" mode  in the QPO data following the 2004 giant flare
of SGR 1806-20 actually seems to be accompanied by a conglomerate of
many peaks in the power spectrum, ranging from about 78 to 105 hz
[\cite{str06}, \cite{wat06}],  implying that at it most extreme
value $\frac{\omega _{lm}^{(1)}}{\omega _{l}^{(0)}}\sim 0.25$. The
92 hertz QPO is attributed to the l=7 toroidal mode [\cite{str06},
\cite{wat06}]. There also seems to be a significant component at
about 80 hz [panels 1-4, and 17 of figure 9 in \cite{str06}], which
could be the l=6 mode. [If the field is axisymmetric and has mirror
symmetry around the equator, then the mechanism for luminosity
variation proposed by \cite{tim07} works for odd l modes, where the
two foot points of a given magnetic field line move in opposite
directions. This introduces a twist in the magnetic field line that
implies a current perturbation.  In the more likely situation that
the field lacks this high degree of symmetry, even l modes are also
possible.] However, there are also modes in the 80-90 hz range and
also significantly above 92, up to 100 hz or more, so it is hard to
interpret the data unambiguously. While there is some evidence for
systematic increase of the frequency with time, the QPO signal in
any given time frame appears to be constant, and in some cases
several bands appear simultaneously, separated by several hz. We
conclude that if the splitting is due to a  magnetic field, it is at
least 2 percent, and at most about 25 percent.

 According to the calculations here, the
variation is expected to be of order 0.4 to
$0.5\frac{B_{0}^{2}}{8\pi \mu }$, as m ranges from 0 to 7. This
implies that $\frac{B_{0}^{2}}{8\pi \mu }\sim  0.04$, or $B\sim 0.2
\sqrt{8\pi \mu }$
if we interpret the 2
percent splitting to be due to magnetic effects.

If the entire range of from 78 to 105 hz is attributed to magnetic
splitting, our first order approximation is only marginal at this
strength, and a higher order calculation would give a slightly lower
value for the frequency shift by a given field strength, so all we
can say is that the field is of the same order as $(8\pi\mu)^{1/2}$.

The value of $(4\pi\mu)^{1/2}$ has been estimated by  to be about $6
\times 10^{15}$ Gauss (\cite{Thom95}), so, if we assume a magnetic
splitting of 2 percent, the value of B appears to be of order $1.7
\times 10^{15}$ Gauss or higher, in reasonably good agreement with
that estimated for the dipole field component, $1.6 \times 10^{15}$
(\cite{wood02,palm05}). Note that in the geometry used here, the
field lines do not particularly lie within the crust but rather cut
through it vertically at angle $\pi/2 - \theta$, where $\theta$ is
the latitude, so a purely toroidal field would be estimated to be
somewhat weaker than the above estimate by a factor of 30 percent or
so. On the other hand, a toroidal field could easily be somewhat
larger than the poloidal field without affecting the dipole moment.

One possible observational consequence of magnetic splitting of
frequency degeneracy for toroidal oscillations is the fact that it
leaves m and -m modes degenerate to the extent that the field is
axisymmetric about the magnetic axis. In contrast to  the magnetic
splitting between different $|m|$ modes (which appears to be on the
order of several hertz and implies that the beat periods would be
less than the rotation period), the m and -m modes would beat more
slowly, and their beating could possibly be observed on a timescale
of perhaps seconds to tens of seconds. The appearance of a
particular frequency band beating on and off during the QPO activity
could be a signature of the  m and -m modes having [or, more
precisely, of their symmetric and antisymmetric combinations having]
slightly different frequencies. Such beating would be a measure of a
non-axisymmetric component to the field. If, for example, the field
is larger at $\phi = 0$ than at $\phi = \pi/2$, then the symmetric
combination of the m=1 and m=-1 modes, proportional to cos$\phi$,
would have a slightly higher frequency than the anti-symmetric mode,
proportional to sin$\phi$.

The residual m,-m degeneracy could also be removed by rotation of a
neutron star. For toroidal modes, the angular frequency shift due to
rotation in a rotating reference frame attached to the star is $m
\Omega/l(l+1)$ (see eg. \cite{pek61,str91}), where $\Omega$ is the
angular frequency of rotation. With period of rotation $P_{rot}
\simeq 7.5s$ one can see that the splitting is quite small and
modulations of the crustal displacement amplitude (beats)  are
possible with period $T_{mod}=P_{rot}l(l+1)/2m$. Since the physics
of QPO variations in magnetar luminosity is somehow determined by
crustal oscillations [e.g. \cite{tim07}], one might also expect time
modulations in the observed QPO components on time scales of order
30 seconds at $l=m=7$.

Another effect that could be looked for is the different damping
rates of the symmetric and antisymmetric combinations due to
coupling with the core. In general, such damping occurs when the
frequency of the mode matches the resonant Alfven frequency of some
connecting field lines in the core. But the damping rate would
depend on the amplitude of the crustal oscillation at the latitude
and longitudes where the frequency match happens to take place, and
this amplitude can differ among the various linear combinations of
symmetric modes. This suggests that QPO components could resolve to
narrower frequency bands as the more rapidly damped linear
combination gives way to the surviving combination. Detecting such
an effect, however, would require good frequency resolution.

The differences between beating, resolution to the longest lived of
several modes, and continuous frequency drift should all be made
clear. We are unable to see how magnetic splitting or damping by the
continuum leads to continuous frequency drift, and we see little if
any evidence for it in the data of \cite{str06}.

The potential wealth of data available in magnetar seismology awaits
confirmation of an accepted model for it.  Future observations of
intermediate flares, which may be more frequent than giant flares,
may provide badly needed additional data.

\acknowledgements {The authors thank Drs. A. Watts and Y. Lyubarsky
for helpful discussions. We acknowledge the Israel-U.S. Binational
Science Foundation, the Israel Science Foundation, and the Joan and
Robert Arnow chair of Theoretical Astrophysics for generous
support.}

\appendix
\section{Vector spherical harmonics}
Vector spherical harmonics are defined as
\[
\vec{Y}_{JM}^{L}\left( \theta ,\varphi \right) =\sum_{m,\sigma } %
C_{Lm1\sigma }^{JM}Y_{Lm}\left( \theta ,\varphi \right) \vec{e}_{\sigma }
\]
where
\[
\vec{e}_{+1}=-\frac{1}{\sqrt{2}}\left( \vec{e}_{x}+i\ \vec{e}_{y}\right) ,\
\vec{e}_{0}=\vec{e}_{z}\ ,\ \vec{e}_{-1}=\frac{1}{\sqrt{2}}\left( \vec{e}%
_{x}-i\ \vec{e}_{y}\right)
\]
Here L can have values $L=J, J\pm 1$ for a given J .

During the calculations following formulae prove useful (for more
information see e.g. \citep{var}).
\[
\int d\Omega \ \vec{Y}_{J_{1}M_{1}}^{L_{1}\ast }\left( \theta ,\varphi
\right) \cdot \vec{Y}_{J_{2}M_{2}}^{L_{2}}\left( \theta ,\varphi \right)
=\delta _{J_{1}J_{2}}\delta _{L_{1}L_{2}}\delta _{M_{1}M_{2}}
\]

\begin{eqnarray*}
\vec{Y}_{J_{1}M_{1}}^{L_{1}}\left( \theta ,\varphi \right) \times \vec{Y}%
_{J_{2}M_{2}}^{L_{2}}\left( \theta ,\varphi \right)  &=&i\sqrt{\frac{3}{2\pi
}\left( 2J_{1}+1\right) \left( 2J_{2}+1\right) \left( 2L_{1}+1\right) \left(
2L_{2}+1\right) }\cdot  \\
&&\sum_{J,L}\left\{
\begin{array}{ccc}
J_{1} & L_{1} & 1 \\
J_{2} & L_{2} & 1 \\
J & L & 1%
\end{array}%
\right\} C_{L_{1}0L_{2}0}^{L0}C_{J_{1}M_{1}J_{2}M_{2}}^{JM}\vec{Y}_{JM}^{L}\left( \theta
,\varphi \right)
\end{eqnarray*}

\[
\vec{\nabla}\times \left[ f\left( r\right) \vec{Y}_{JM}^{J+1}\left( \theta
,\varphi \right) \right] =i\sqrt{\frac{J}{2J+1}}\left( \frac{d}{dr}+\frac{J+2%
}{r}\right) \ f\left( r\right) \ \vec{Y}_{JM}^{J}\left( \theta ,\varphi
\right)
\]

\begin{eqnarray*}
\vec{\nabla}\times \left[ f\left( r\right) \vec{Y}_{JM}^{J}\left( \theta
,\varphi \right) \right] =i\sqrt{\frac{J}{2J+1}}\left( \frac{d}{dr}-\frac{J}{%
r}\right) \ f\left( r\right) \ \vec{Y}_{JM}^{J+1}\left( \theta ,\varphi
\right)&& \\
+ i \sqrt{\frac{J+1}{2J+1}}\left( \frac{d}{dr}+\frac{J+1}{r}\right) \
f\left( r\right) \ \vec{Y}_{JM}^{J-1}\left( \theta ,\varphi \right)&&
\end{eqnarray*}

\[
\vec{\nabla}\times \left[ f\left( r\right) \vec{Y}_{JM}^{J-1}\left( \theta
,\varphi \right) \right] =i\sqrt{\frac{J+1}{2J+1}}\left( \frac{d}{dr}-\frac{%
J-1}{r}\right) \ f\left( r\right) \ \vec{Y}_{JM}^{J}\left( \theta ,\varphi
\right)
\]
Now, taking into account that $\vec{u}_{lm}^{(0)}$ and $\vec{e}_{z}$
can be represented as  [e.g. \citep{var}]
\[
\vec{u}_{lm}^{(0)}=-i\sqrt{l(l+1)} w_{l}^{(0)}\left( r\right) \vec{Y}_{lm}^{l} , \quad
\vec{e}_{z}=\sqrt{4\pi }\, \vec{Y}_{10}^{0}
\]
one can use the above formulae to get the result
(\ref{a1},\ref{b1}).

\clearpage
\begin{deluxetable}{cccc}
\tabletypesize{\scriptsize} \tablecaption{Values of the function $F(l,m)$ for $l=5,7,9$ }
\label{ff} \tablewidth{0pt}
\tablehead{\colhead{m} & \colhead{$F(5,m)$} &
\colhead{$F(7,m)$} & \colhead{$F(9,m)$}} \startdata
0 & 0.487   & 0.493   & 0.496 \\
1 & 0.472   & 0.485   & 0.49  \\
2 & 0.426   & 0.459   & 0.474 \\
3 & 0.349   & 0.416   & 0.447 \\
4 & 0.241   & 0.356   & 0.409 \\
5 & 0.103   & 0.279   & 0.36  \\
6 & \nodata & 0.184   & 0.3   \\
7 & \nodata & 0.074   & 0.23  \\
8 & \nodata & \nodata & 0.149 \\
9 & \nodata & \nodata & 0.057 \\
\enddata
\end{deluxetable}

\end{document}